\documentclass[11pt]{article}

\usepackage[in]{fullpage}
\usepackage[colorlinks=true,citecolor=blue,linkcolor=black]{hyperref}
\usepackage{graphicx}
\usepackage{amsmath}
\usepackage{mathtools}
\usepackage{blkarray}
\newcommand{\matindex}[1]{\mbox{\scriptsize#1}}
\usepackage{amssymb}
\usepackage{booktabs}
\usepackage{multicol}
\usepackage{float}

\usepackage[ruled,vlined]{algorithm2e}

\newenvironment{protocol}[1][htb]
  {\renewcommand{\algorithmcfname}{Protocol}
   \begin{algorithm}[#1]%
  }{\end{algorithm}}

\usepackage{mathtools}

\usepackage{parskip}
\begin{document}

\title{A Graph Partitioning Algorithm for Leak Detection in \\ Water Distribution Networks}
\author{Aravind Rajeswaran, Sridharakumar Narasimhan, Shankar Narasimhan}
\date{Systems \& Control Group \\[0.05in] Indian Institute of Technology Madras \\[0.1in]
\normalsize aravindr@smail.iitm.ac.in, $\lbrace$ sridharkrn, naras $\rbrace$@iitm.ac.in }

\maketitle

\begin{abstract}
Leak detection in urban water distribution networks~(WDNs) is challenging given their scale, complexity, and limited instrumentation. We present a technique for leak detection in WDNs, which involves making additional flow measurements on-demand, and repeated use of water balance. Graph partitioning is used to determine the location of flow measurements, with the objective to minimize the measurement cost. We follow a multi-stage divide and conquer approach. In every stage, a section of the WDN identified to contain the leak is partitioned into two or more sub-networks, and water balance is used to trace the leak to one of these sub-networks. This process is recursively continued until the desired resolution is achieved. We investigate different methods for solving the arising graph partitioning problem like integer linear programming (ILP) and spectral bisection. The proposed methods are tested on large scale benchmark networks, and our results indicate that on average, less than 3\% of the pipes need to be measured for finding the leak in large networks.
\end{abstract}

\section{Introduction}
The problem of leak detection in Water Distribution Networks (WDNs) is of significant importance for effective management and water quality control~\cite{CA02,PR10}. Leaky distribution systems are inefficient due to water loss, energy wastage, and unreliable water quality: especially in case of underground leaks. These effects are even more pronounced in urban centers of developing countries where the networks are poorly instrumented and maintained.

In case of WDNs, leaks or losses are quantified using unaccounted-for water (UFW). High levels of UFW are detrimental to financial viability of the system. Losses in WDNs are a combined effect of real losses like leaks in pipes or joints, as well as other means like water thefts and unauthorized consumption~\cite{GG11}. Given the growing concern towards uncertainty in quality water supplies, the problem of leak detection and control has grown in importance. Various techniques based on acoustic methods and magnetic flux leakage~\cite{MW01,ZS11,AC09} are available to determine the location of defect (either small defect like corrosion, or large leaks) in a single pipe. However, these methods could be time consuming, expensive, or disruptive in nature. Thus, it is beneficial to use these techniques after narrowing down the leak to a small part of the network.

One approach to leak detection involves the use of hydraulic models and simulators. Available measurements are used to estimate the location of a leak which match the sensor measurements closely.  This method is generally called inverse analysis~\cite{Chen94} and requires solving a large optimization problem.
In order to use this approach, measurements of flow rates and pressures at a large number of intermediate locations are required, in addition to source pressure and demand flows. In well instrumented networks, some flow and pressure sensors are installed for the purpose of District Metered Area (DMA) sectorization, but these are few in number. A more severe limitation of pressure-reading based methods is that predictions depend on precise estimates of model parameters like pipe friction factors, which are difficult to obtain. Practical applicability of this method to large scale networks have proven to be a hard task, as reported by some researchers~\cite{Stephens04,Stephens05}.

\begin{figure}[b!]
\begin{center}
\begin{multicols}{2}
	\includegraphics[width=0.45\textwidth]{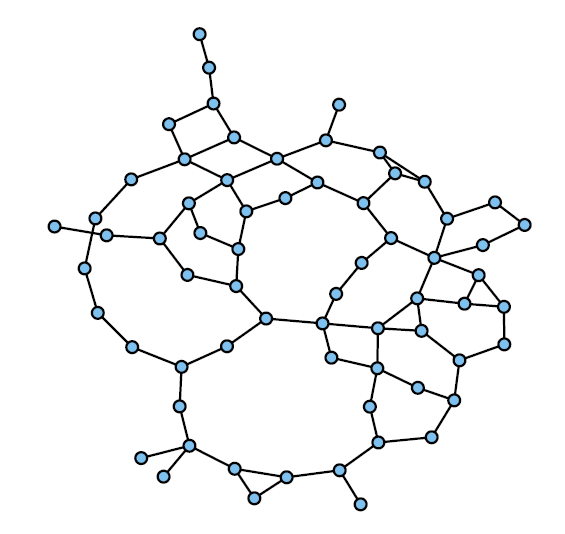} \\
	(a) Network with a leaky node
	
	\includegraphics[width=0.45\textwidth]{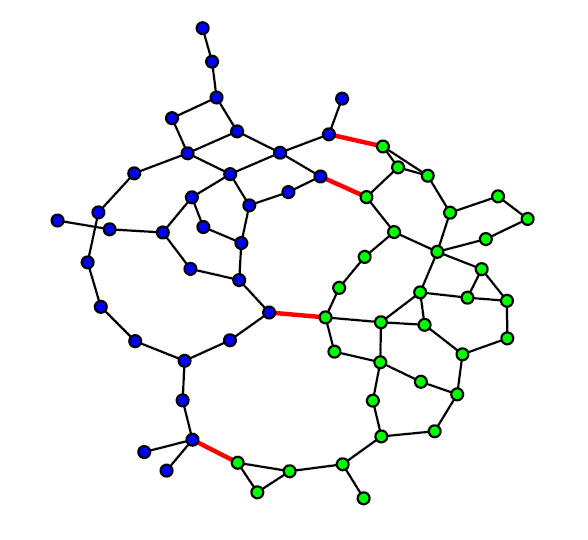} \\
	(b) Querying a set of pipes
\end{multicols}
\end{center}
\caption{Illustration of querying the edges for flows and identifying the leaky part of the network.}
\end{figure}

In order to overcome the above difficulties, and to explore a new line of research, we propose a method for leak detection which uses only flow measurements that are repeatedly performed on-demand in field campaigns. We call this process of obtaining flow measurement in a pipe (on-demand) as \textbf{querying} the pipe for flow. Further, since the only property of leak we exploit is loss of material~(water), the method is equally applicable to any form of loss including thefts - which is not the case for hydraulic model based methods. Even though we show results of our method on WDNs, the method itself is much more general and pertains to any distribution system obeying conservation laws.

\newpage
To briefly illustrate the idea, consider the network shown in Fig.~1(a). Let us say that some node in this network is leaky, and our objective is to find it. By querying the edges in red in Fig.~1(b), we can trace the leak to either of the two parts of the network (shown in blue and green). This is possible by exploiting water balance (or conservation laws in general) as will be shown in subsequent sections. By performing this operation repeatedly, we can arrive at a small part of the network which contains the leak.

Querying a pipe requires access to it, which may be buried underground at a depth of about two meters. Hence, there is a non-negligible cost associated with every query. Therefore, it is important to minimize queries (or query cost), which requires a strategic field campaign. 
An ideal field campaign should possess the following characteristics: (i)~it should be systematic and arise out of a clear objective; (ii)~it should scale to large sectors or the whole network in absence of DMAs; (iii)~must be capable of assimilating information from other sources (like existing sensors); (iv)~should be optimal, requiring only few queries. An algorithmic solution to development of such a field campaign is the subject of this paper.

\section{Review of Algebraic Graph Theory}
Before the formal problem setup, we briefly review the basics of algebraic graph theory relevant to this work. Specifically, we review the representation of WDNs as graphs and matrices, and survey relevant properties. See chapter~7 of \cite{Deo} for further discussion.

\noindent {\bf Definition 2.1} A graph $(\mathbf{G})$ is a tuple $\mathbf{G}(\mathbf{N},\mathbf{E})$ comprising the set of vertices $\mathbf{N}$ and edges $\mathbf{E}$ which are 2-element subsets of $\mathbf{N}$. The number of vertices and edges in the graph are $n$ and $m$ respectively. The graph could be directed or undirected. We use the following terms interchangably to suit the particular context: graph and network; vertices and nodes; edges, links, and pipes.

The nodes of the network can be classified as source nodes where water is fed into the network, demand nodes or sink nodes where water is removed from the network for supplying to the consumers, and transmission nodes which aid in redistributing the flows. The edges of the network represent the pipes of the WDN. We choose an {\it undirected graph} representation for the network. However, we associate a sign convention with each edge to help identify the direction of flow. Flow will be negative if it is in the opposite direction to the chosen sign.

\noindent {\bf Definition 2.2} The adjacency matrix is defined by the relationship:  $A_{ij} = 1$ if nodes i and j are connected by a pipe and 0 otherwise.

\noindent {\bf Definition 2.3} The directed incidence matrix $\mathbf{J}$ is defined by the relationship:
$$ J_{ik} = 
\begin{cases}
+1 & \text{if edge k connects nodes i and j, and }  i<j \\
-1 & \text{if edge k connects nodes i and j, and }  i>j \\
0 & \text{if edge k is not incident on node i}
\end{cases}
$$
The sign convention for $\mathbf{J}$ can in fact be chosen arbitrarily and the above assignment is only one particular choice. 

\noindent {\bf Definition 2.4} The degree of node $i$, is the number of edges incident on the node and denoted by $\mathrm{deg}(i)$. The degree matrix $\mathbf{D}$ is a diagonal matrix containing the degree of each node along the diagonal entries, i.e., $D_{ii}=deg(i)$ and $D_{ij}=0, i\neq j$.

\noindent {\bf Definition 2.5} The Laplacian ($\mathbf{L}$) of a graph is defined by the relationship $\mathbf{L} = \mathbf{D-A}$, where $\mathbf{D}$ and $\mathbf{A}$ are the degree and adjacency matrices, respectively.

The adjacency and incidence matrices characterize the network completely. The other matrices can be computed with their knowledge. We also review some useful properties of these matrices.

\noindent {\bf Property 2.1} The Laplacian matrix is positive semi-definite.

\noindent {\bf Property 2.2} The smallest eigenvalue of the Laplacian matrix is 0. The vector $\mathbf{v}=[1,1,\ldots,1]^T$ (or simply $\mathbf{v} = \mathbf{1}$) satisfies $\mathbf{Lv}=\mathbf{0}$ and hence is an eigenvector correpsonding to the 0 eigenvalue and  belongs to the nullspace of $L$.

\noindent {\bf Property 2.3} The number of times 0 appears as an eigenvalue of the Laplacian (both algebraic and geometric multiplicity) is the number of connected components in the graph.

\noindent {\bf Proposition} The Laplacian matrix is identically equal to the positive semi-definite matrix $\mathbf{JJ^T}$

\noindent {\bf Proof}\\
Define $\mathbf{Z}=\mathbf{JJ^T - L}$ with eigenvalues $\lambda_1,\ldots,\lambda_n$ and corresponding eigenvectors $\mathbf{v_1},\ldots,\mathbf{v_n}$. From direct verification, $\mathbf{x^T JJ^T x} = \sum_{(i,j) \in \mathbf{E}} (x_i - x_j)^2 = \mathbf{x^T L x} \ \forall \mathbf{x}$. Hence, $\mathbf{x^T Z x} =0$ $\forall \mathbf{x}$ and in particular holds true for the eigenvectors, i.e., $\mathbf{v_i^T Z v_i} = \lambda_i \mathbf{v_i^T v_i} = 0$. This implies that all eigenvalues $\lambda_i$ are 0 and hence $\mathbf{Z=0}$ implying $\mathbf{JJ^T = L}$

\noindent {\bf Definition 2.6} A subgraph $\mathbf{S(N_S,E_S)}$ is formed from a graph $\mathbf{G(N,E)}$ such that $\mathbf{N_S} \subseteq \mathbf{N}$ and $\mathbf{E_S}$ contains all the edges with both endpoints in $\mathbf{N_S}$.

\noindent {\bf Definition 2.7} A partition of $\mathbf{G(N,E)}$ consists of two subgraphs $\mathbf{S}$ and $\mathbf{\bar{S}}$ such that $\mathbf{N_{\bar{S}}} = \mathbf{N} \setminus \mathbf{N_S}$

\noindent {\bf Definition 2.8} The cut-set of partition $( \mathbf{S}, \mathbf{\bar{S}} )$ is the set of all edges having one endpoint in $\mathbf{N_S}$ and the other in $\mathbf{N_{\bar{S}}}$. 
Formally, $\text{cut}( \mathbf{S}, \mathbf{\bar{S}} ) = \mathbf{E} \setminus \left( \mathbf{E_S} \cup \mathbf{E_{\bar{S}}} \right) $

\noindent {\bf Definition 2.9} If each edge is associated with a cost, then the cut-cost of partition $( \mathbf{S}, \mathbf{\bar{S}} )$ is the sum of costs of each edge present in the cut-set. We denote this with $R( \mathbf{S}, \mathbf{\bar{S}} )$

\noindent {\bf Definition 2.10} Graph partitioning problem: Find a partition that minimizes $R( \mathbf{S}, \mathbf{\bar{S}} )$, while satisfying certain constraints (typically cardinality constraints on $\mathbf{N_S}$). A partition is said to be balanced if the number of nodes in the partitions are approximately equal. 

\section{Problem Formulation}
In this section, we first present the problem statement in the most general form. Next, present a protocol which describes our formulation and solution procedure. Finally, we also present an example to illustrate both the problem and the protocol.

\newpage
The general objective of leak detection is to locate the (or all) leaky unit (pipe or junction) using the information that is available about the network. Invariably, additional information in the form of some flow or pressure readings; and system parameters like effective dimension and friction factor of pipes would be required. Hence, most leak detection procedures address in some form, the trade-off between accuracy or confidence of identified leak and collection of additional information. Due to difficulties in obtaining some of these parameters (like friction factor) and other reasons specified in Section~I, we explore a procedure which requires the use of only flow measurements which are queried on demand. We first present the assumptions made in the formulation.
\begin{enumerate}
    \itemsep0em
    \item The WDN is in steady state condition.
    \item The topology of the WDN (ie the graph representation) is known. 
    \item We possess portable flow meters which can detect the flow rate as well as direction. One example is ultrasonic flow meters which use time of flight principle.
\end{enumerate}
Additionally, we make the following assumptions to simplify and aid the presentation. We will later show simple methods to avoid them.
\begin{enumerate}
    \itemsep0em
    \item All supply (source) and demand (consumption) flow rates are measured continuously. No other permanent sensors are available. We later present techniques to incorporate any additional sensors that may be available.
    \item There is only one leak in the network, and it is present in a node. The algorithm naturally extends to multiple leaks and leaks along any point of the pipe as well.
    \item The sensors measurements are noiseless. If the measurements contain random errors or bias, we can overcome this by using various statistical techniques for decision making~\cite{Naras15,Rajes15}.
\end{enumerate}

With these assumptions, we now present the protocol. The goal is to find a partitioning algorithm for the protocol that will minimize the Cost. \\
\begin{protocol}[H]
\SetAlgoLined
{\bf Input:} Graph $\mathbf{L(N,E)}$ containing leaky node, $\delta$ (threshold) \\
{\bf Initialize:} $ \mathbf{G} \leftarrow \mathbf{L} $; 
Cost $\leftarrow 0$ \\
\While{ $\text{size}(\mathbf{G}) > \delta$ }{
    $(\mathbf{S},\mathbf{\bar{S}}) \leftarrow $ partition $(\mathbf{G})$ \\
    $\mathbf{G} \leftarrow $ find leaky partition $(\mathbf{S},\mathbf{\bar{S}})$ \\
    Cost $\leftarrow$ Cost + $R(\mathbf{S},\mathbf{\bar{S}})$
}
{\bf Result:} Leaky node is in vertex set of $\mathbf{G}$
\caption{Leak detection procedure}
\end{protocol}
Each step in the loop will be presented in greater detail in subsequent sections. We devote two sections (4 and 5) for development of the partitioning algorithm. Section 3.1 provides an example to illustrate the idea. Section 3.2 outlines the procedure to find the leaky partition using water balance. Finally, we present some extensions to the protocol in Section~8 and appendix. 

\subsection{Illustrative Example}

\begin{figure}[b!]
\begin{center}
\includegraphics[width=\textwidth]{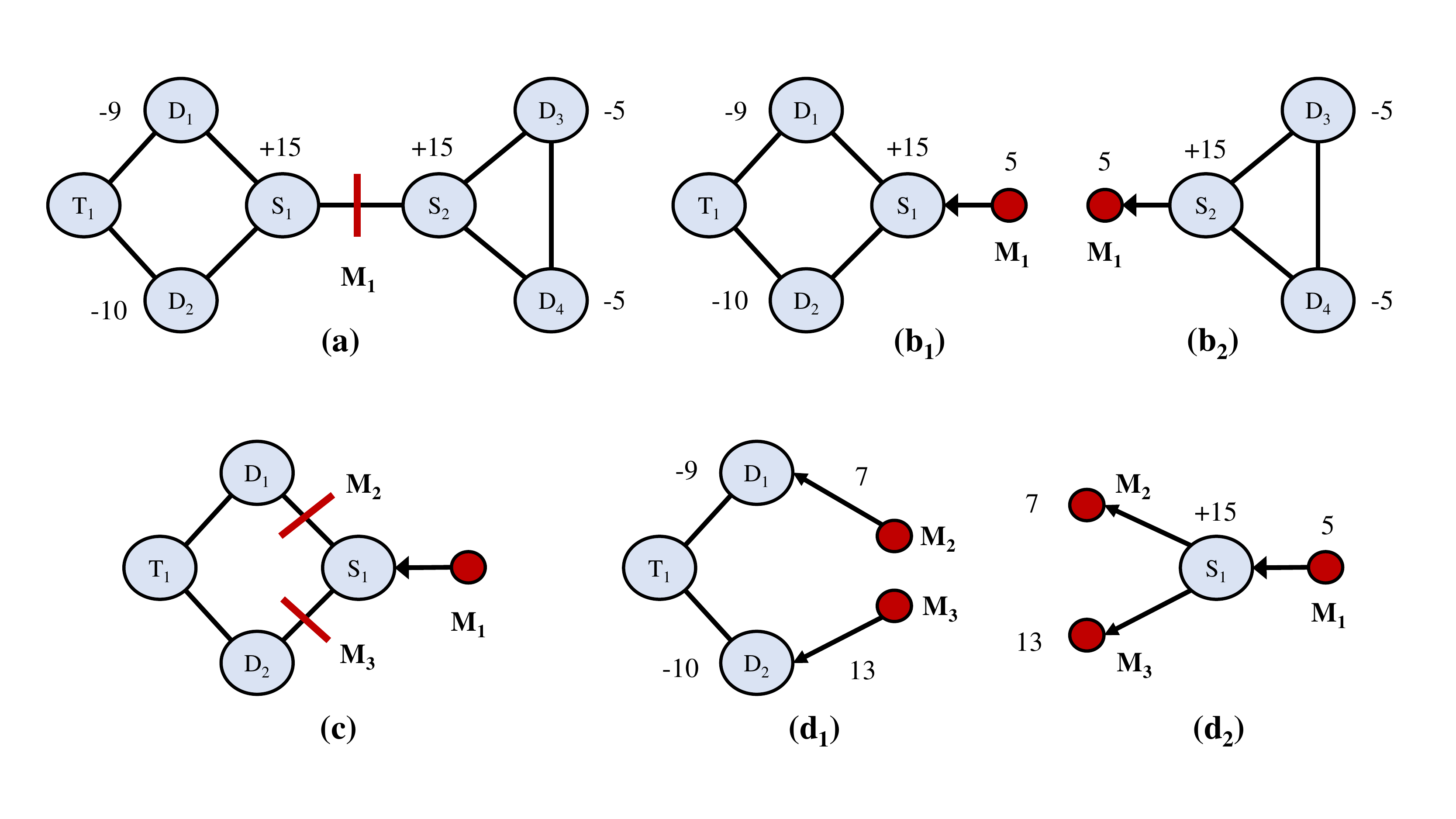}
\caption{Illustrative example on a simple flow network: $S_i$, $D_i$, $T_i$ represent source, demand, and transmission nodes respectively. Production and consumption rates are also shown.} 
\label{fig:seq}
\end{center}
\end{figure}

Consider the WDN shown in Fig.~\ref{fig:seq}(a) where nodes $S_1, S_2$ are supply nodes, $D_1,D_2,D_3,D_4$ are  demand nodes, and $T_1$ is a transmission node. The demand and supply rates are measured and the values are presented next to the respective nodes (with signs). The different figures represent the information that is uncovered about the network through the querying process. From Fig.~2(a), we can see that a total of 30 flow units are supplied (at $S_1$ and $S_2$), whereas only 29 units are consumed at the demand nodes ($D_1$ to $D_4$). Since the network is in steady state, this is possible only when $1$ unit of flow is removed from the system in the form of a leak. Thus, a water balance reveals the presence of a leak, and our objective is to find the location.

It is clearly not possible to identify the leak based on Fig.~2(a), but only ascertain that there is one. We make a query of measurement ($M_1$) as shown, which produces the two partitions shown in Fig.~2($b_1$) and Fig.~2($b_2$). A similar balance indicates that the leak is present in the sub-graph comprised of the nodes $\lbrace S_1, T_1, D_1, D_2 \rbrace$. Note that we must add the additional source term of $M_1$ to the subgraph in order to account for the external flows to or from this subgraph. We proceed further by querying additional edges for flows as shown in Fig.~2($d_1$) and Fig.~2($d_2$). This approach can be mathematically formalized and presented as an optimization problem with the objective to minimize the query cost in the protocol presented earlier.

\subsection{Water balance for identifying leaks}

\begin{figure}[h!]
\begin{center}
\includegraphics[width=0.6\textwidth]{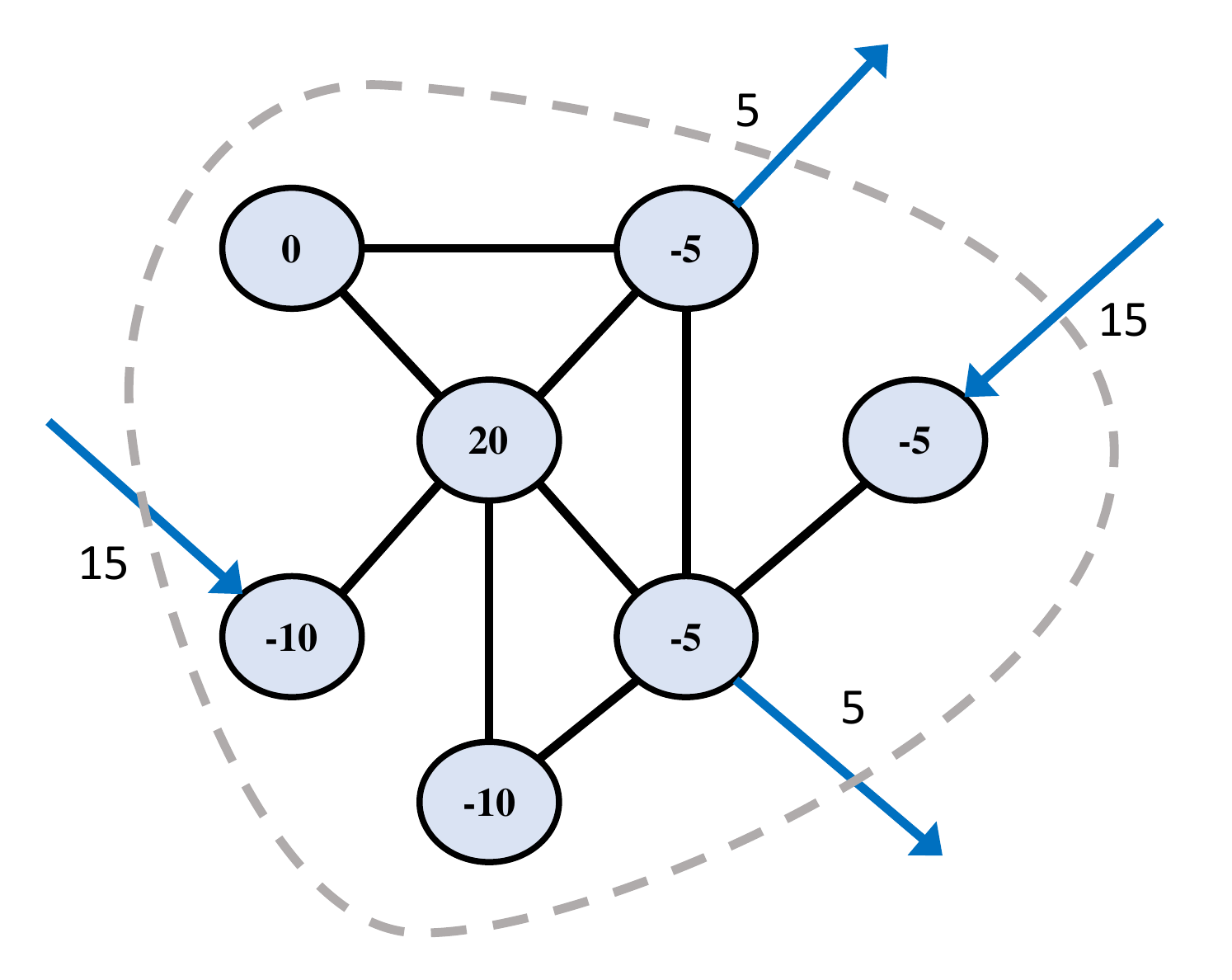}
\caption{Envelope around a sub-network of source, sink, and transmission nodes} \label{fig:envelope}
\end{center}
\end{figure}

Water balance is a special case of mass conservation. When considering a generic envelope encompassing a set of nodes and edges, it is possible to perform a water balance around this envelope given knowledge of source and sink terms within the envelope (assumed known) and flow rates in the pipes crossing the envelope boundary (which are to be measured/queried). This idea is illustrated in Fig.~\ref{fig:envelope}. The generalized balance equation under steady state condition takes the form:
$$ \text{in + production = out + consumption} $$
Hence knowledge of flow direction is required to correctly apply the balance equation, for which we need appropriate instrumentation. If such instrumentation is not available, alternate methods to ascertain flow direction must be used. One approach could be to use nominal case hydraulic simulations, and assume the flow directions do not change in presence of leak. This is likely valid only if the magnitude of leak is small. In general, we assume that either appropriate instrumentation or alternate methods to find flow directions are available. For the envelope shown in Fig.~\ref{fig:envelope}, we see that in=30, production=20, out=10, and consumption=35. Hence the balance is violated, and 5 units of flow are unaccounted for. Thus, the leak can be traced to this small part of a larger WDN.

\newpage

\section{Graph Partitioning - ILP formulation}
In this section we motivate the graph partitioning procedure and provide various Integer Linear Programming (ILP) formulations for the same. As per Protocol~1, we require a partitioning algorithm that will minimize the Cost. In practice, finding a partitioning algorithm or policy for minimizing the cumulative Cost is difficult. This is because for any partition, we will not know which subgraph contains the leak till the partition is actually made and the corresponding cost incurred. Thus, the number of possibilities which we need to search over is exponentially large. To overcome this, we provide an approximate approach or heuristic where at each iteration in the protocol loop, we minimize only $R( \mathbf{S}, \mathbf{\bar{S}} )$ subject to some constraints which indirectly take into account costs incurred in subsequent iterations.

We define an indicator variable $\mathbf{x} \in \lbrace 0,1 \rbrace^n$ that indicates the partition to which a node belongs. Formally,  $x_i$ is defined as follows: $x_i=1$  implies node $i \in \mathbf{S}$  and $x_i=0$  implies node $i \in \mathbf{\bar{S}}$. We restrict our attention to the non-trivial case of a connected graph. Consider the $k^\mathrm{th}$ element of $\mathbf{J^T x}$, where edge $k$ connects nodes $i$ and $j$.
$$ \sum_{l=1}^{l=n} J_{lk} \ x_{l} = J_{ik} x_i + J_{jk} x_j $$
This is because only two rows of the $k^\mathrm{th}$ column of $\mathbf{J}$ are non zero (by construction). 
Further, assuming without loss of generality that $i < j$,
$$ J_{ik} x_i + J_{jk} x_j  = x_i - x_j $$ 
We now make the following observations:
$$\sum_{l=1}^{l=n} J_{lk} x_{l} = x_i - x_j = 
\begin{cases}
\ \ 0 & x_i = 1, x_j = 1 \ \text{or} \ x_i = 0, x_j = 0 \\
\ \ 1 & x_i = 1 \ \& \ x_j = 0 \\ 
-1 & x_i = 0 \ \& \ x_j = 1 
\end{cases}$$
The cut-size is the number of non-zero entries in the vector $\mathbf{J^Tx}$. This is related to $R(\mathbf{S},\mathbf{\bar{S}})$ as:
\begin{equation}
R(\mathbf{S},\mathbf{\bar{S}}) = \sum_{k=1}^{k=m} \ w_k \left| \ \sum_{l=1}^{l=n} J_{lk} \ x_{l} \ \right| 
\end{equation}
Where the cost of querying edge $k$ is $w_k$. When presenting the results, we set $w_k = 1 \forall k$, but it is clear that the method naturally extends to arbitrary values.
As a proxy for minimizing the Cost, we propose to minimize $R(\mathbf{S},\mathbf{\bar{S}})$ while ensuring that the partitions are balanced. This constraint is important to avoid partitions that myopically reduce $R(\mathbf{S},\mathbf{\bar{S}})$ to provide lopsided partitions which may endure large cut-costs in subsequent iterations. Parallels can be drawn with binary search, except that in our problem cost for different splits are different. 
The two objectives of minimizing $R(\mathbf{S},\mathbf{\bar{S}})$ and keeping the partition balanced are likely to be conflicting. Thus it is natural to pose the problem generally as a multi-objective optimization problem.
\begin{equation}
\begin{aligned}
& \underset{\mathbf{x}}{\text{min.}}
& & \left| n - 2 \times \sum_{i=1}^{i=n} x_i \right| \text{\ \ and \ \ } \sum_{k=1}^{k=m} \ w_k \left| \ \sum_{l=1}^{l=n} J_{lk} \ x_{l} \ \right|  \\
\\
& \text{s.t.}
& & \mathbf{x} \in \lbrace 0,1 \rbrace^n
\end{aligned}
\end{equation}
We first propose to remove the absolute expressions so that we can formulate the problem as a standard ILP. Consider the constraint: 
$$ 2 \times \sum_{i=1}^{i=n} x_i \leq n $$
This serves dual purposes. Firstly, it removes the absolute value expression in the first objective. Additionally, it helps in pruning the search space by removing symmetries. For any feasible solution $\mathbf{x}$ the solution $\mathbf{1 - x}$ is equally valid and would produce the same value for the objective function. By introducing the above constraint, this symmetry is broken.
In order to remove the absolute value expression in the second objective, we introduce two new m-vectors ($\mathbf{t_1}$ and $\mathbf{t_2}$) as decision variables  such that:
$$ \mathbf{t_1 - t_2} = \mathbf{J^T \ x} $$
$$ \mathbf{t_1 + t_2} \leq \mathbf{1} $$
$$ \mathbf{t_1} \in [ 0,1 ]^m \ \ \ \mathbf{t_2} \in [ 0,1 ]^m $$

Minimizing the second objective is now equivalent to minimizing  $\sum_{k=1}^{m} t_1(k) + t_2(k)$, since an element of the vectors, $t_1(k)$ or $t_2(k)$ take the value $1$ only when $x_i - x_j = \pm 1$, and are forced to $0$ (minimization) whenever $x_i - x_j = 0$. Hence the optimization problem can now be written as:

\begin{equation}
\label{eq:original_formulation}
\begin{aligned}
& \underset{\mathbf{x,t_1,t_2}}{\text{min.}}
& & \overset{\mathrm{Size\ Disparity}}{\overbrace{ 
	 	\left( -2 \times \mathbf{1^T x} \right) }} \hspace{5pt} \text{ and } \hspace{5pt}
	\overset{\mathrm{Cut-Cost}}{\overbrace{    
	 	\left( \mathbf{w^T t_1 + w^T t_2} \right) }} \\
& \text{s.t.}
& & \mathbf{t_1 - t_2} = \mathbf{J^T \ x} \\
& {}
& & \mathbf{t_1 + t_2} \leq \mathbf{1} \\
& {}
& & \mathbf{1^T x} \leq 0.5 n \\
& {}
& & \mathbf{x} \in \lbrace 0,1 \rbrace^n \ \ \ \mathbf{t_1} \in [ 0,1 ]^m \ \ \ \mathbf{t_2} \in [ 0,1 ]^m
\end{aligned}
\end{equation}
Since the problem formulation is that of a multi objective optimization problem, there are many different ways to solve it. Each method is based on some notion of the relative importance of the different objective terms. One simple method is to scalarize the objective function by assigning relative weights to the different objective terms. In this case however, there is no obvious metric to trade off one for the other. One suggested method in literature is normalized cuts~\cite{Ncuts}, which proposes a ratio measure. However, in our problem, the end goal is to minimize Cost, for which the n-cut scalarization doesn't have a clear physical interpretation. We study two different paradigms which provide better physical insight for this application.

\subsection{Lexicographic solution}
The first method is that of a lexicographic optimization where the idea is that one objective is infinitely more important than the other. Lexicographic solutions to multi objective optimization has been extensively researched~\cite{Sherali1983} and also successfully applied to sensor placement problems~\cite{Bhushan2002,Bhushan2008}. Here the first objective is given precedence and is minimized first and among multiple solutions which can achieve this, the one which minimizes the second objective is picked, and so on. In this problem, we give more importance to the balanced partitioning objective, and admit only those solutions which produce perfectly balanced partitions. Conceptually, among the $n \choose n/2$ possible solutions for obtaining balanced partitions, that solution which minimizes the cut-cost is chosen. However, we do not explicitly enumerate all these possibilities, but instead solve the following ILP.

\begin{equation}
\label{eq:lexicographic}
\begin{aligned}
& \underset{\mathbf{x,t_1,t_2}}{\text{min.}}
& & \mathbf{w^T t_1 + w^T t_2}
\\
& \text{s.t.}
& & \mathbf{t_1 - t_2} = \mathbf{J^T \ x} \\
& {}
& & \mathbf{t_1 + t_2} \leq \mathbf{1} \\
& {}
& & \sum_{i=1}^{i=n} x_i = \lfloor \frac{n}{2} \rfloor \\
& {}
& & \mathbf{x} \in \lbrace 0,1 \rbrace^n \ \ \ \mathbf{t_1} \in [ 0,1 ]^m \ \ \ \mathbf{t_2} \in [ 0,1 ]^m
\end{aligned}
\end{equation}

\subsection{Goal programming}

The second method is goal programming, where we set a nominal goal for one objective. For example, this can be introduced in the form of a constraint, so that the search space is confined to those situations which meet this goal. We use this idea for our problem to get a good handle over the partition sizes. By deviating a little from exact bisection, we may be able to reduce the cut-size significantly. In such cases goal programming can be very effective. We assign a goal on the partition size to take the form $\sum_{i=1}^{n} x_i \geq (0.5-\gamma)n$ which guarantees a minimum size for both partitions. $\gamma$ is a parameter which defines the level of goal. Based on our simulations, a good choice is $\gamma = 0.1$, so that partitions have a minimum size of $0.4n$.

In (5), we have added an additional term to the objective function, $\frac{\epsilon}{n} \mathbf{1^T x}$. By choosing $\epsilon$ appropriately, we can have \mbox{$\frac{\epsilon}{n} \mathbf{1^T x} < min(\mathbf{w})$} and hence less than the minimum possible change in $( \mathbf{w^T t_1 + w^T t_2} )$. Thus, addition of this term cannot alter the optimal value of cut-cost. The purpose of this term is to ensure that if there are multiple minimum cut-cost solutions which meet the partition size goal, we would obtain the most balanced partition.
\newpage
\begin{equation}
\label{eq:goal_programming}
\begin{aligned}
& \underset{\mathbf{x,t_1,t_2}}{\text{min.}}
& & \mathbf{w^T t_1 + w^T t_2} + \left( -\frac{\epsilon}{n} \mathbf{1^T x} \right)
\\
& \text{s.t.}
& & \mathbf{t_1 - t_2} = \mathbf{J^T \ x} \\
& {}
& & \mathbf{t_1 + t_2} \leq \mathbf{1} \\
& {}
& & \sum_{i=1}^{i=n} x_i \leq\lfloor \frac{n}{2} \rfloor \\
& {}
& & \sum_{i=1}^{i=n} x_i \geq \lceil (\frac{1}{2}-\gamma)n \rceil \\
& {}
& & \mathbf{x} \in \lbrace 0,1 \rbrace^n \ \ \ \mathbf{t_1} \in [ 0,1 ]^m \ \ \ \mathbf{t_2} \in [ 0,1 ]^m
\end{aligned}
\end{equation}

\section{Graph Partitioning - Approximation}
The general problem of graph partitioning with partition size or cardinality constraints are NP hard~\cite{SA09}. Hence, the ILP models formulated in previous sections must be solved directly to obtain the optimal solutions. For large problems, this may not be feasible, and hence we discus some approximate solution methods.

The idea of approximation algorithms for ILPs involve two steps: relaxing some constraints to solve a simpler problem, and a rounding-off step where solutions consistent with the actual constraints are recovered from the relaxed solutions. Graph partitioning has many approximation algorithms in literature which have been successfully used in different domains. For our application, approximation algorithms have two uses: it can be used to find quick and reliable estimates of upper-bound associated with the field campaign, thereby help make informed policy calls regarding the feasibility of the field campaign. Additionally, it can also be used for the first few levels of very large networks, where ILPs become computationally expensive. For the subsequent levels, when network size has reduced greatly, the ILP algorithm can be used.

The relaxation step is both problem and application specific. For example, if one clearly knows the sizes of the partition - a scenario common in circuit design where number of components to be placed on a chip is known, a popular method of choice is the Kernighan and Lin algorithm~\cite{KL70}. This method is not applicable to our problem since cut-cost is directly tied to our overall objective, and partition sizes cannot be accurately predicted. Under such circumstances, methods from spectral graph theory are more appropriate. Our approach in spirit follows from the goal programming ILP formulation, and we ultimately arrive at a result which is similar to the spectral bisection method~\cite{PSL90} but with subtle differences and alternative interpretations.

Consider an assignment variable for nodes to different partitions chosen as $\mathbf{z} \in \lbrace -1,1 \rbrace^n$ which is equivalent to the earlier choice of $\mathbf{x} \in \lbrace 0,1 \rbrace^n$ through the transformation $\mathbf{z} = (2\mathbf{x}- \mathbf{1})$. We again wish to arrive at an assignment that minimizes cut-cost subject to partition size constraints.
$$ R(\mathbf{S},\mathbf{\bar{S}}) = \frac{1}{2} \|\mathbf{J^Tz}\|_1  =  \frac{1}{4}  \|\mathbf{J^Tz}\|_2^2 $$ 
where $\|.\|_1$ and $\|.\|_2$ represent the 1-norm and 2-norm respectively. For cases where edges have different costs, we can use the simple modification: $\lbrace J_{ik} = w_k, J_{jk} = -w_k \rbrace \forall k$ such that edge $k$ connects nodes $i$ and $j$. 
In the previous section, the objective was formulated using the 1-norm. We now choose to minimize the 2-norm to obtain an approximate analytical solution. Assigning some relative cost $\mu$ (unknown) to the two objectives, the problem can be posed as:
\begin{equation}
\begin{aligned}
& \underset{z}{\text{min.}}
& & \underset{\mathrm{Cut-Cost}}{\underbrace{ 
	 	\left( \mathbf{ z^T J J^T z } \right)}} +
	\underset{\mathrm{Size\ Disparity}}{\underbrace{    
	 	\mu \left( \mathbf{1^T z} \right)^2}} \\
& {\text{s.t.}}
& & \mathbf{z} \in \lbrace -1,1 \rbrace^n 
\end{aligned}
\end{equation}
The goal programming size constraint will be imposed explicitly at a later stage. We relax the integral constraint on $\mathbf{z}$, viz., $\mathbf{z} \in \lbrace -1,1 \rbrace^n $ to $\mathbf{z} \in \mathbb{R}^n$ and $\mathbf{z^Tz}=n$. Next we express $\mathbf{z}$ using the orthonormal set of eigenvectors of $\mathbf{J J^T}$. Let $\lambda_1,\lambda_2,\ldots,\lambda_n$ be the eigenvalues of $\mathbf{J J^T}$ sorted in ascending order with eigenvectors  $\mathbf{u_1,u_2,}\ldots,\mathbf{u_n}$ respectively. Defining
$\mathbf{U} = [\mathbf{u_1, u_2}, \ldots, \mathbf{u_n}]$
we can write $\mathbf{z} = \mathbf{U \alpha}$ where $\mathbf{\alpha}$ is the vector of projections onto $\mathbf{u_1,u_2,}\ldots,\mathbf{u_n}$.
From Property 3.2, we have $\lambda_1 = 0$, $\mathbf{u_1} = \frac{\mathbf{1}}{\sqrt{n}}$ and $\lambda_i>0, ~i=2,\ldots,n$. The constraint $\mathbf{z^Tz}=n$ is equivalent to  
$ \alpha_1^2 + \alpha_2^2 + ... + \alpha_n^2 = n $
With this change of variables, the optimization problem in (6) after relaxation becomes:

\begin{equation}
\begin{aligned}
& \underset{\alpha}{\text{min.}}
& & \alpha_2^2 \lambda_2 + \alpha_3^2 \lambda_3 ... + \alpha_n^2 \lambda_n  + n \mu \alpha_1^2 \\
& \text{s.t.}
& & \alpha_1^2 + \alpha_2^2 + ... + \alpha_n^2 = n \\
& {}
& {}
\end{aligned}
\end{equation}
\noindent with  $\lambda_2 \leq \lambda_3 ... \leq \lambda_n$. 
The   above problem can be solved analytically as follows:
\begin{enumerate}
\itemsep-0.5em
\item If $n \mu \leq \lambda_2$, then $\alpha_1^2 = n$ and $\alpha_i = 0$ $\forall i \neq 1$ 
\item If $n \mu > \lambda_2$, then $\alpha_2^2 = n$ and $\alpha_i = 0$ $\forall i \neq 2$ 
\end{enumerate}
The first solution indicates that if cut-cost is significantly more than cost associated with size disparity in partitions, the obvious solution is to not partition at all. This solution is trivial and is discarded. The second solution indicates that if cost associated with disparity is more than a certain threshold, then the solution is to partition such that $\alpha_2^2 = n$. This suggests the assignment choice as $\mathbf{z} = \sqrt{n} \mathbf{u_2}$ where $\mathbf{u_2}$ is the eigenvector corresponding to the second smallest eigenvalue, also known as the Fiedler vector. 
Since $\mathbf{u_2}\neq \mathbf{0}$ is orthogonal to $\mathbf{u_1} = \frac{\mathbf{1}}{\sqrt{n}}$, $\mathbf{u_2}$ is non-trivial.
In order to obtain an integer solution,  we employ a simple round off procedure to obtain the solution $\mathbf{z}$ that is consistent with problem specifications, and also maximizes $\alpha_2^2$. The final solution is: 

$$\mathbf{z} = \text{sgn}(\mathbf{u_2})$$
Note that the above solution maximizes $\alpha_2^2$ which is only an approximation of the original problem. In order to minimize the true problem~(6), we need to consider the relative magnitudes of the different eigenvalues which is possible only in a combinatorial setting. In fact, it is this approximation that enables us to arrive at a computationally tractable solution.

\begin{figure}[b!]
\begin{center}
\includegraphics[width = 0.6\textwidth]{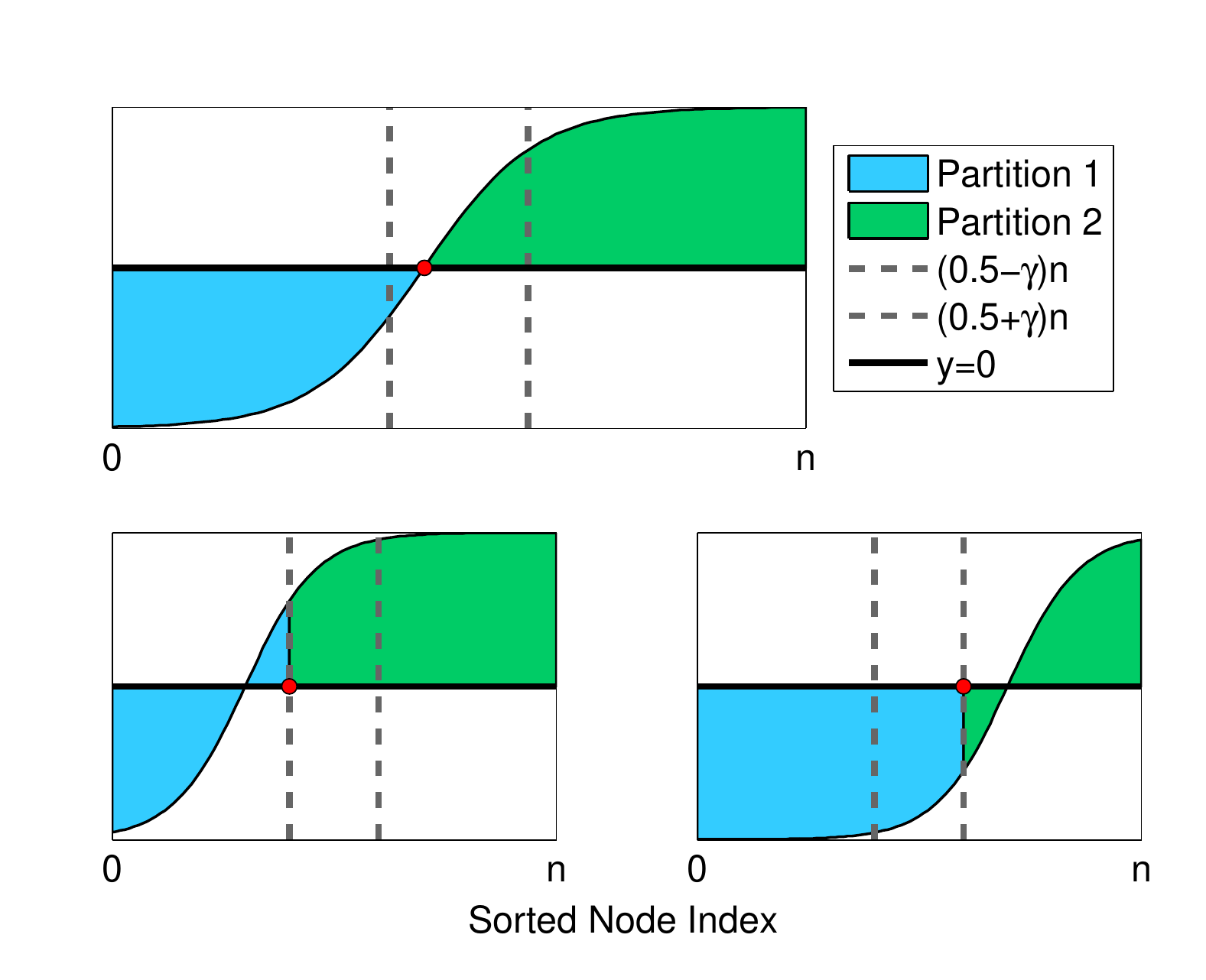}
\caption{Incorporating size restriction to approximation algorithm.} \label{fig:approx_alg}
\end{center}
\end{figure}

Partitioning based on entries of Fiedler vector is known by the name of spectral bisection~\cite{PSL90} and is known to produce skewed partitions~\cite{Ncuts}. This problem can be tackled by explicitly imposing a goal programming constraint as shown in Figure~4. We sort the entries of $\mathbf{u_2}$ in ascending order, and normally assign partitions based on sign of the entry corresponding to each node. If we get skewed partitions, we can cut-off the partitions at the threshold defined by the minimum partition sizes. This is shown schematically in Fig.\ref{fig:approx_alg}. This assignment ensures that $\alpha_2^2$ is maximized when adhering to the partition size constraints. This is because there is a fixed number of sign mismatches that would occur between $z_i$ and $u_2(i)$ which reduces the value of $\alpha_2$ from its maximum possible value. By sorting and assigning nodes to partitions such that sign mismatches always occur with $u_2 (i)$ of least magnitude, the maximum possible value of $\alpha_2$ is achieved in presence of the partition size constraint.

{\bf Remark:} While we have presented two methods here (ILP and approximation scheme) which work well for the target application as seen through case studies, researchers have attempted other approximation algorithms, and the field of graph partitioning is very rich in literature. Some of these methods employ the use of semi-definite programming and randomized algorithms~\cite{VG11, SA09}. 
We do not present the results of these algorithms since the size of benchmark networks considered in this paper were not large enough to render ILPs computationally infeasible, and the spectral bisection method provides adequate performance. However, if necessary, it is trivial to incorporate other approximate partitioning methods into Protocol~1.

\section{Results and Discussion}
To test the proposed methods, we have chosen representative water distribution networks used frequently in literature. These include the EXNET, Richmond, DTown, and Colorado Springs networks. Researchers~\cite{Yazdani2011} have studied the topology of these networks, with emphasis on analyzing properties like link density, clustering coefficient, betweenness centrality etc.

\begin{figure}[H]
\begin{center}
\begin{multicols}{2}
	\includegraphics[width = 15pc]{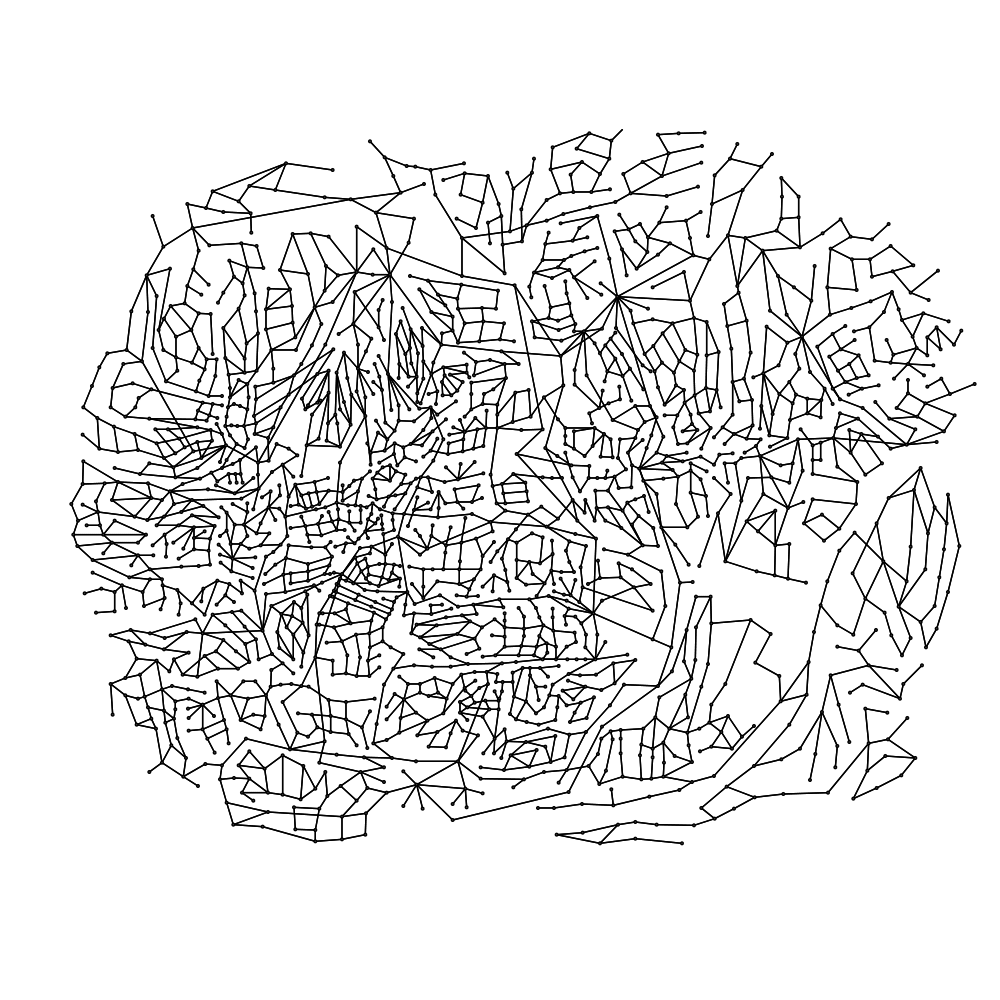} \\
	(a) Exnet network \\
	
	\includegraphics[width = 15pc]{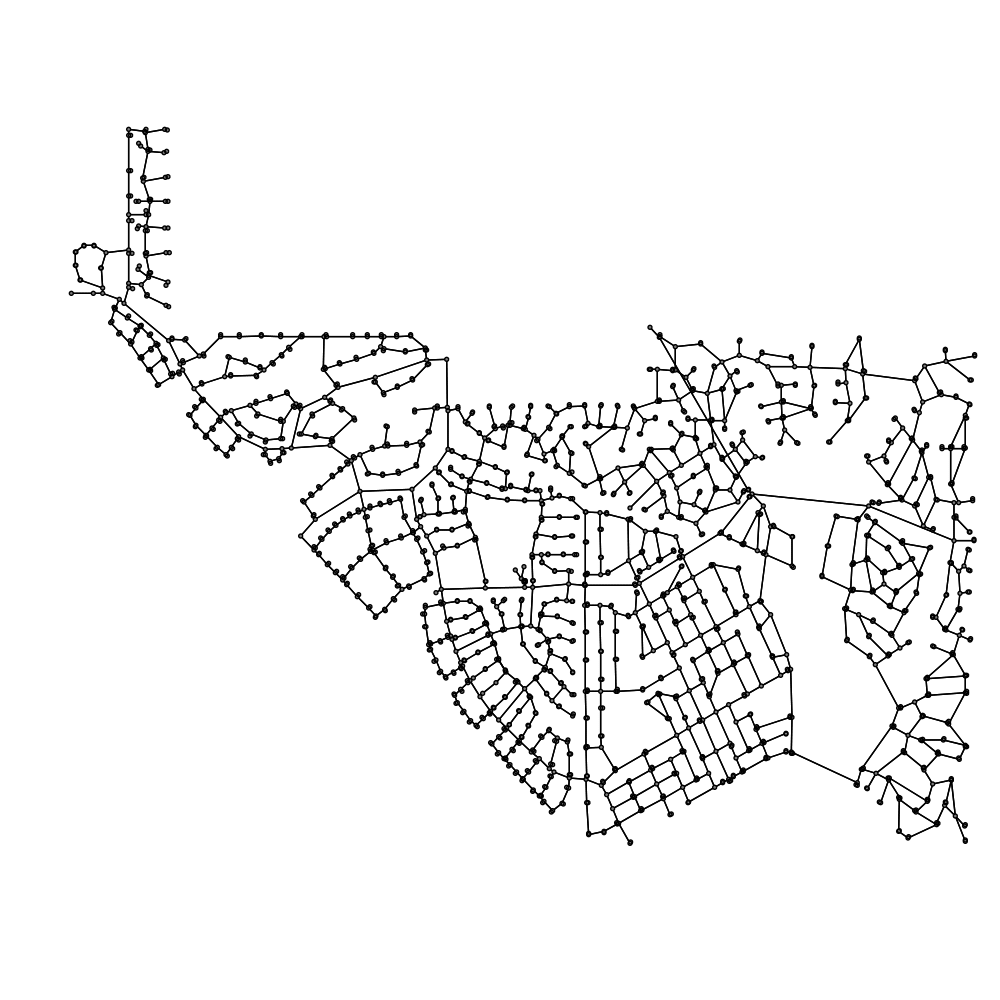} \\
	(b) Colorado Springs \\
	
	\includegraphics[width = 15pc]{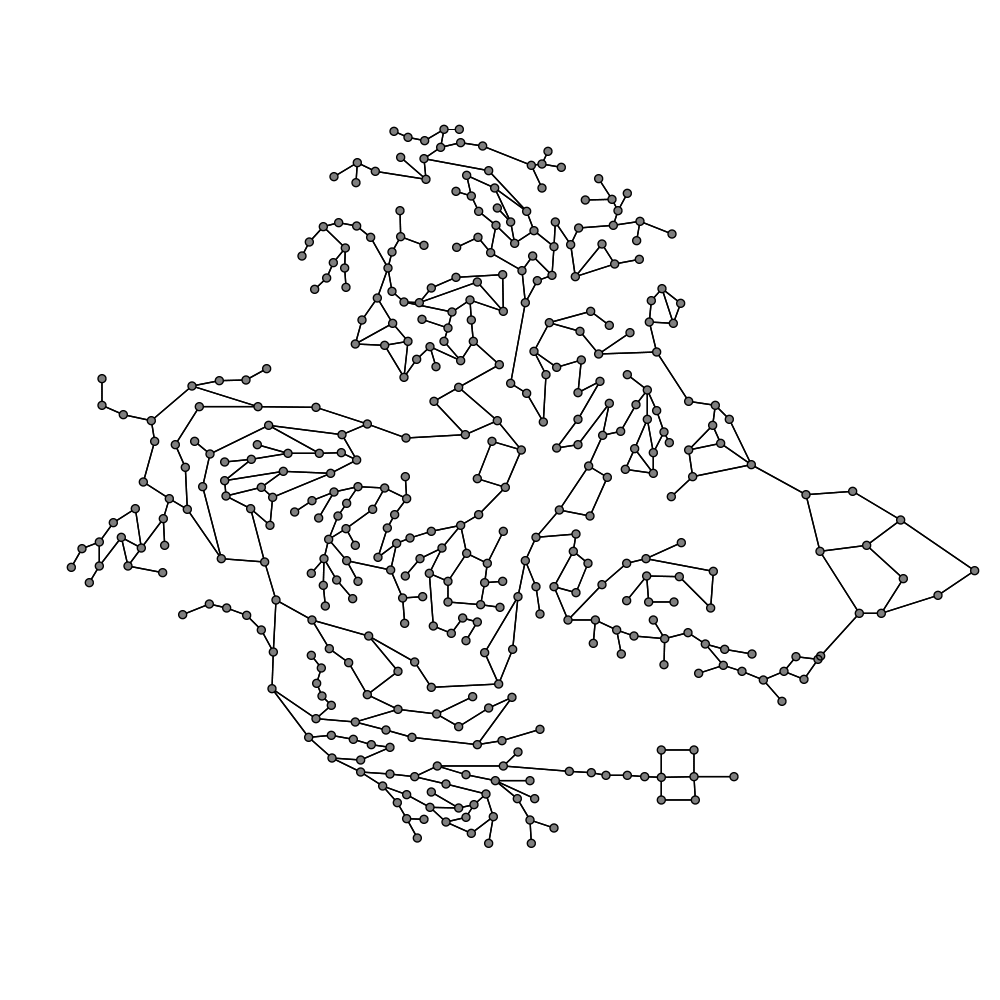} \\
	(c) DTown network \\
	
	\includegraphics[width = 15pc]{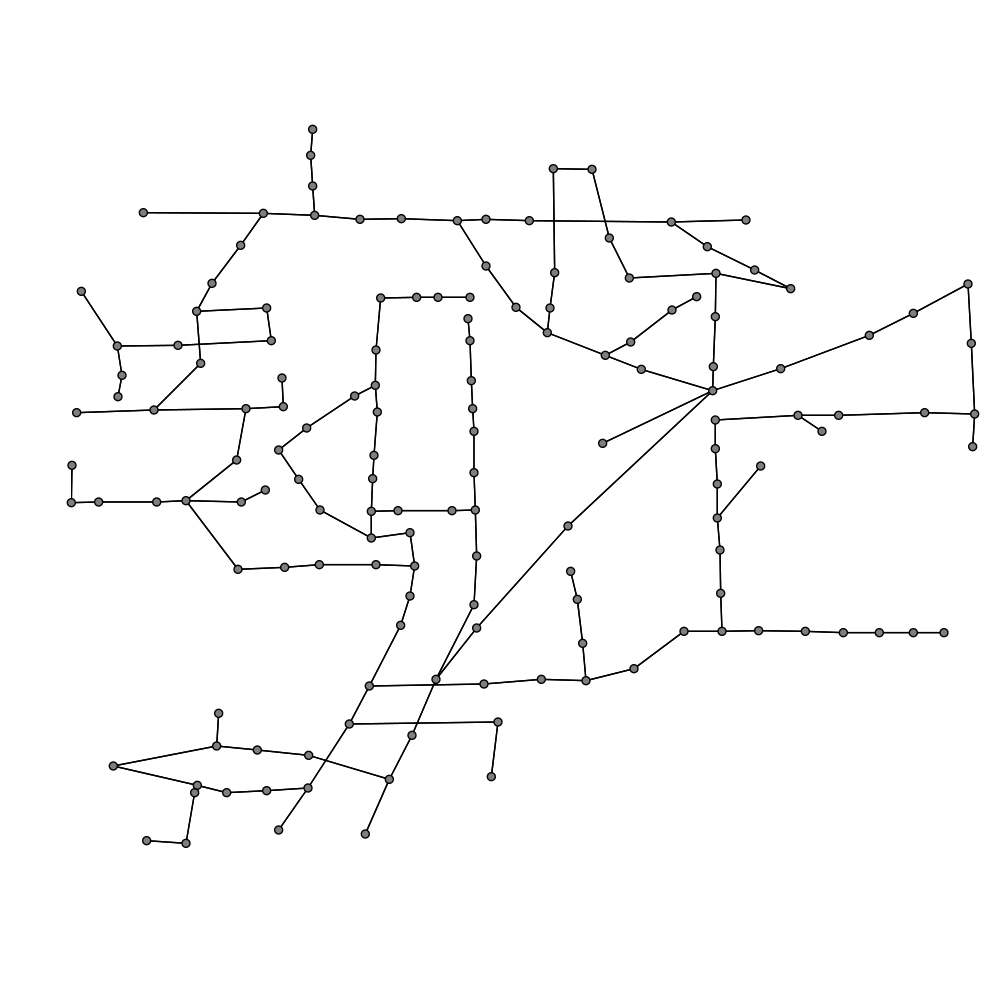} \\
	(d) Bangalore network \\
\end{multicols}
\end{center}
\caption{Layouts of the different networks considered in this paper. Network structure and layout was obtained from the website of The Centre for Water Systems at the University of Exeter.} \label{fig:network_layouts}
\end{figure}

We have chosen these networks due to the wide spectrum of size, formation, and organizational patterns; and hence representative of most WDNs~\cite{Yazdani2011}. The EXNET network is a large realistic benchmark problem used for multi-objective optimization of water systems. The Colorado Springs network is an example with multiple water supply sources, while the Richmond network is a sub-network of the Yorkshire Water system in the UK with a single reservoir. The DTown network was used in the Battle of the Water Network II (BWN-II) as a design problem. In addition, we have also tested the algorithm on one sector of the Bangalore water distribution network, which is smaller in size compared to the other ``full" networks, to study how the methods perform at smaller scales. Some important properties of these networks are summarized in Table~1. The layouts of these networks are illustrated in Fig.~\ref{fig:network_layouts}

\begin{table}[H]
\begin{center}
\caption{Properties of the networks studied. ($n$ and $m$ are the number of nodes and edges respectively; q is the link density ($\frac{2m}{n(n-1)}$); $<k>$ and $k_{max}$ are the mean and maximum node degrees)}
\begin{tabular}{@{}lcccccc@{}}
\toprule
\multicolumn{1}{c}{\textbf{Network}} & \textbf{n} & \textbf{m} & \textbf{q} & $\mathbf{<k>}$ & $\mathbf{k_\mathrm{max}}$ \\ \midrule
\textbf{Exnet}                       & 1893       & 2418       & 1.35e-3    & 2.55              & 10              \\
\textbf{CO. Springs}                 & 1786       & 1992       & 1.25e-3    & 2.23              & 4               \\
\textbf{Richmond}                    & 872        & 957        & 2.52e-3    & 2.20              & 4               \\
\textbf{Dtown}                       & 401        & 459        & 5.72e-3    & 2.29              & 5               \\
\textbf{Bangalore}                   & 150        & 155        & 1.43e-2    & 2.07              & 5               \\ \bottomrule
\end{tabular}
\end{center}
\label{tab1:network properties}
\end{table}

\newpage

\subsection{Case Study 1 - Leak in nodes}
In the first case study, we assume that leaks are always present in nodes. This is a continuation of the assumptions made in the problem formulation. We choose to apply the algorithm repeatedly till we find the leaky node (this corresponds to $\delta = 1$ in Protocol 1). We simulate the leak in every node, and apply the protocol repeatedly till we find it, and record the number of queries required. This corresponds to a full enumerative study since we iterate over every possible leak scenario. The summary statistics are tabulated in Tables~2 and 3.

\newpage
For the goal programming formulation, a nominal goal of $\gamma = 0.1$ was chosen. This corresponds to requirement that both partitions have at least $0.4n$ nodes.  As expected, the number of queries required are minimal when solving the goal programming ILP. It is also observed that the approximation algorithm performs reasonably well, with the number of queries much smaller than the size of network (both nodes and edges). As an example, when using the goal programming ILP formulation, for the largest network Exnet, the maximum number of queries required is only $1.7\%$ of the number of edges. For the Bangalore network, which is the smallest considered, about $8.4\%$ of the edges need to be queried in the worst case. This is expected since in small networks or sub-networks, modular features are less prominent. Similar trends are observed when using the approximation algorithm as well. Under worst case scenario, the fraction of queries (against edges) required when using the approximation method is $2.9\%$ for Exnet and $8.4\%$ for Bangalore.

\begin{table}[H]
\begin{center}
\caption{Results of ILP using Goal-Programming method}
\begin{tabular}{@{}lccccc@{}}
\toprule
\multicolumn{1}{c}{{\bf Network}} & \multicolumn{5}{c}{{\bf Number of measurements}}               \\ \midrule
\multicolumn{1}{c}{{\bf }}        & {\bf mean} & {\bf median} & {\bf mode} & {\bf max} & {\bf std} \\
{\bf Exnet}                       & 29.74      & 31           & 34         & 42        & 5.88      \\
{\bf CO. Springs}                 & 23.78      & 22           & 22         & 38        & 4.92      \\
{\bf Richmond}                    & 11.80      & 11           & 10         & 20        & 2.23      \\
{\bf Dtown}                       & 11.10      & 11           & 10         & 16        & 1.60      \\
{\bf Bangalore}                   & 10.44      & 10           & 10         & 13        & 1.10      \\ \bottomrule
\end{tabular}
\end{center}
\label{tab:ILP_GP_Node}
\end{table}

\begin{table}[H]
\begin{center}
\caption{Results of approximation algorithm}
\begin{tabular}{@{}lccccc@{}}
\toprule
\multicolumn{1}{c}{{\bf Network}} & \multicolumn{5}{c}{{\bf Number of measurements}}               \\ \midrule
\multicolumn{1}{c}{{\bf }}        & {\bf mean} & {\bf median} & {\bf mode} & {\bf max} & {\bf std} \\
{\bf Exnet}                       & 54.58      & 53           & 50         & 71        & 6.40      \\
{\bf CO. Springs}                 & 35.90      & 33           & 31         & 51        & 6.81      \\
{\bf Richmond}                    & 13.56      & 13           & 13         & 23        & 3.38      \\
{\bf Dtown}                       & 12.26      & 12           & 12         & 18        & 1.78      \\
{\bf Bangalore}                   & 10.31      & 10           & 9          & 13        & 1.37      \\ \bottomrule
\end{tabular}
\end{center}
\label{tab:Approx_Node}
\end{table}

\subsection{Case Study 2 - Leak in edges/pipes}
In most practical cases, it is unlikely that leaks can be present only in nodes. In general, we would not know upfront the nature of leak (whether it is at a node or on a pipe). In the appendix, we discuss a simple modification of the method to enable leak detection in pipes as well. 

\begin{figure}[H]
\begin{center}
\begin{multicols}{2}
	\includegraphics[width = 0.45\textwidth]{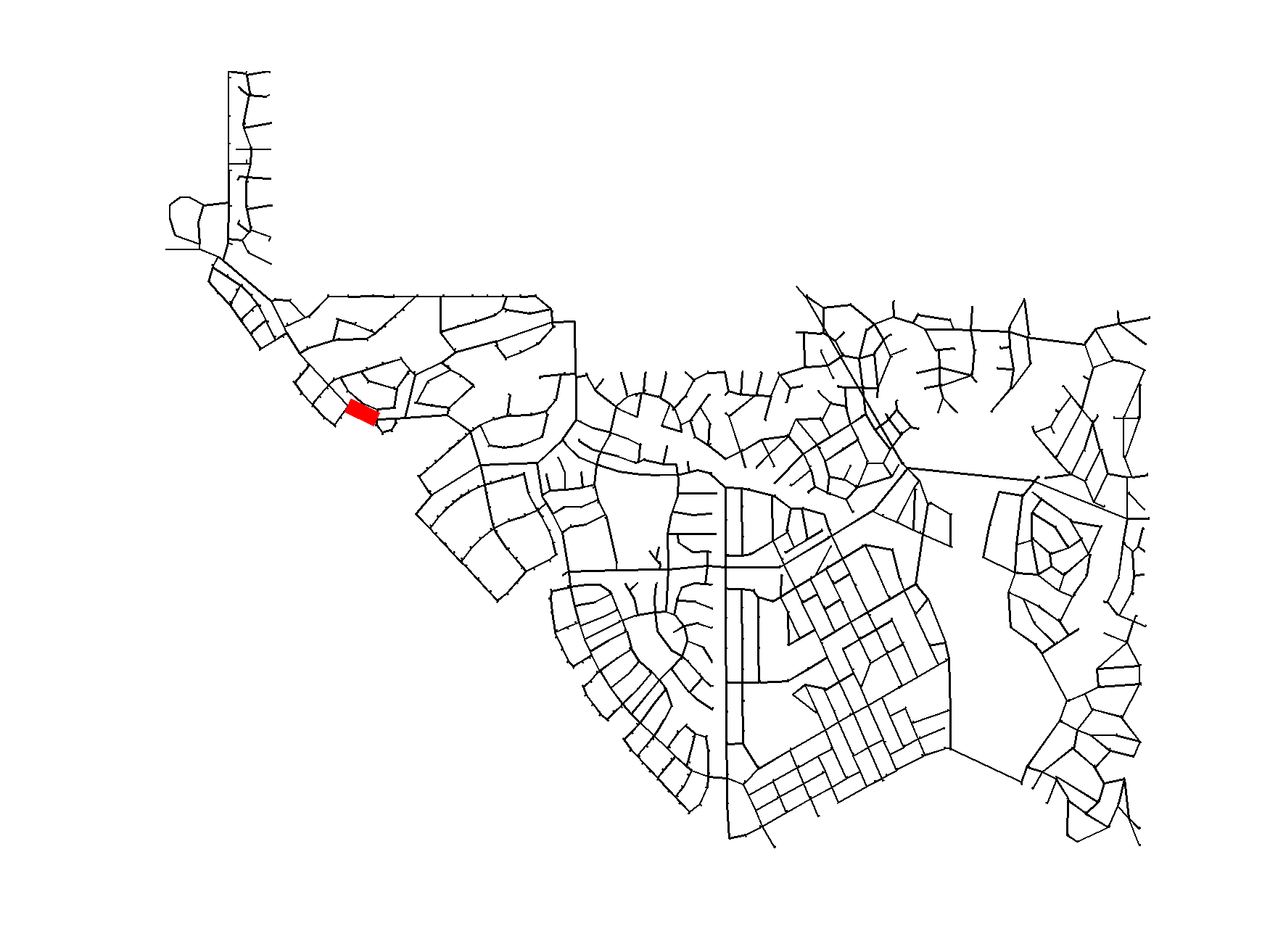} \\
    (a) network with leaky pipe \\
	
	\includegraphics[width = 0.45\textwidth]{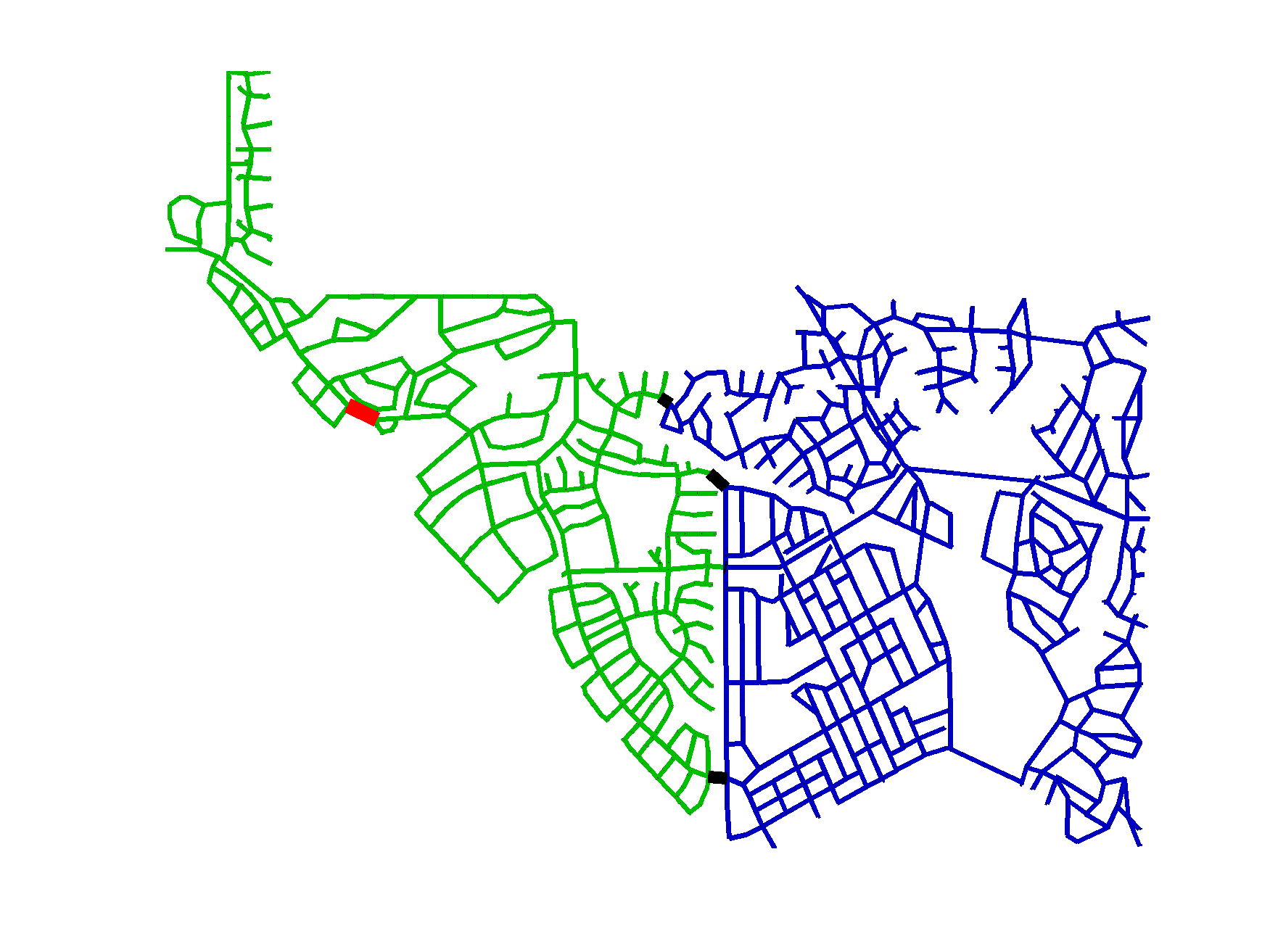} \\
	(b) first set of queries \\
	
	\includegraphics[width = 0.45\textwidth]{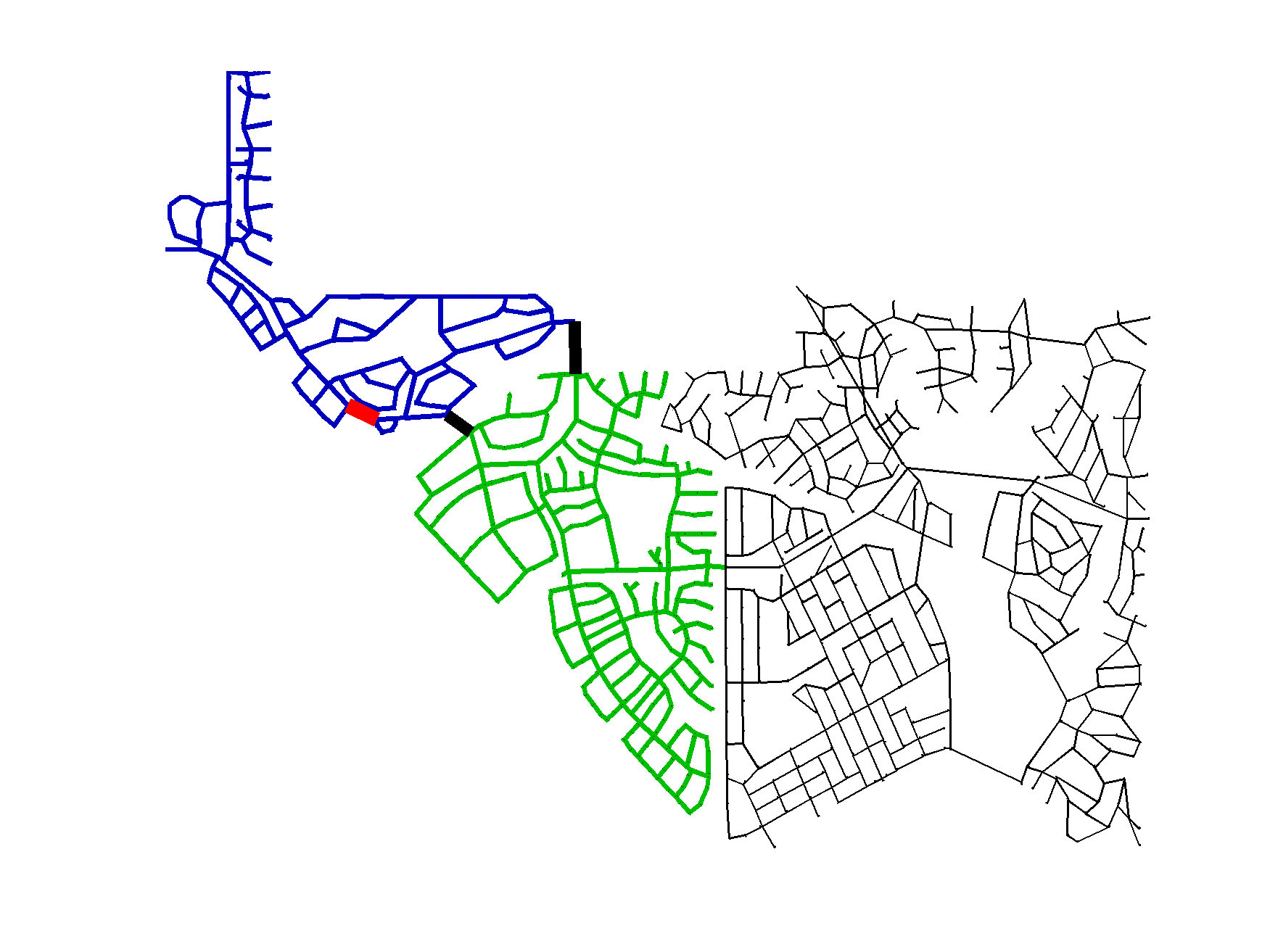} \\
	(c) second set of queries \\
	
	\includegraphics[width = 0.45\textwidth]{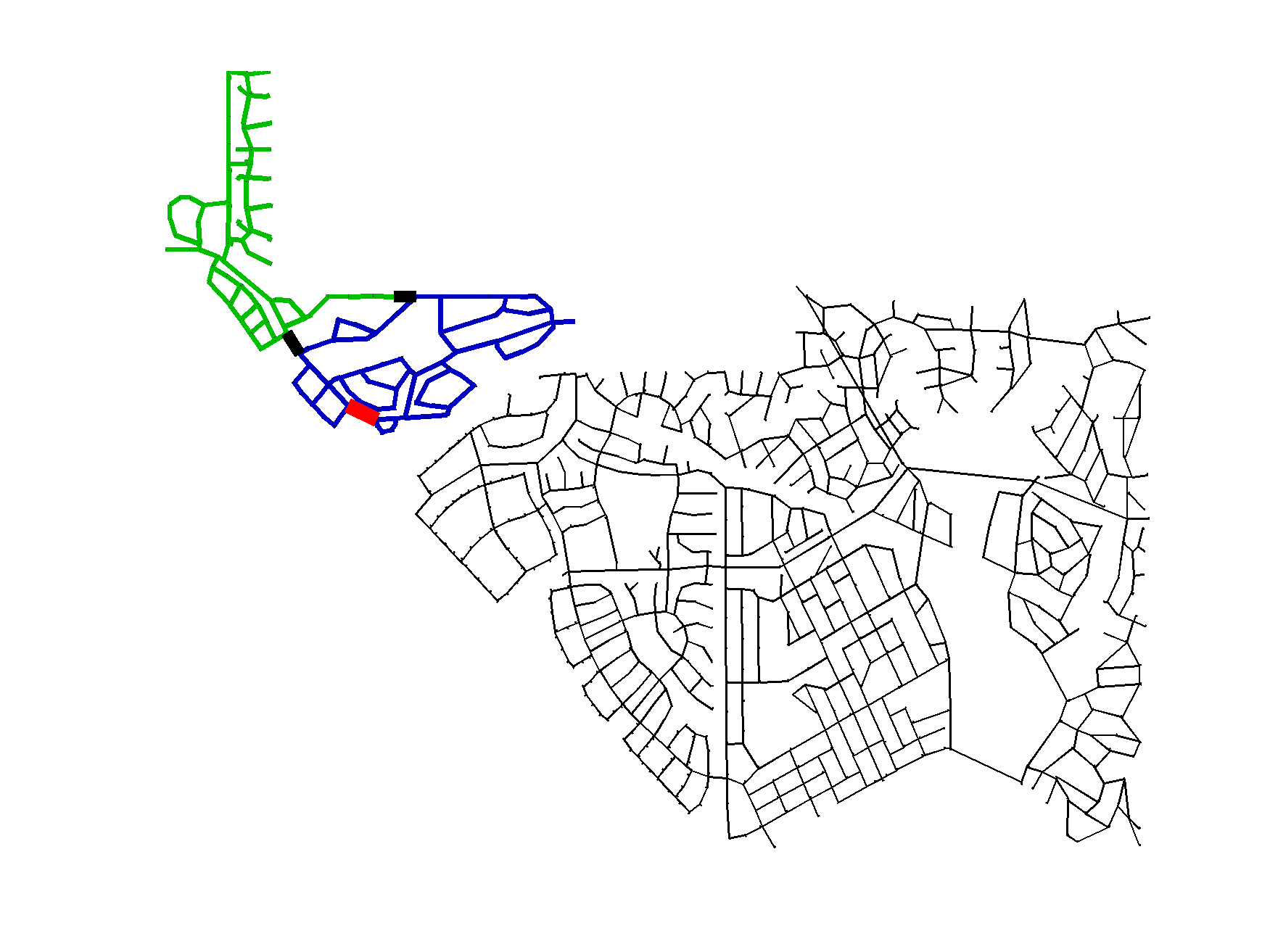} \\
	(d) third set of queries \\
	
\end{multicols}
\end{center}
\caption{Sequence of pipes that are queried to identify the leak shown in (a) in red.}
\label{fig:trace_leak_example}
\end{figure}

For this case study, we again run the protocol till we find the leaky pipe, though it can be stopped prematurely if required. As an illustration consider the situation shown in Fig.~\ref{fig:trace_leak_example}(a) which contains a leaky pipe shown in red. We query different edges in sequence as shown in Fig.~\ref{fig:trace_leak_example}(b,c,d) where the thick black lines indicate pipes that are queried. The process is continued till we converge to the leaky pipe. We again perform a full enumerative study to generate the results presented in Tables~4 and 5.

For goal programming, we set $\gamma = 0.1$. Similar trends to Case Study 1 are observed where the ILP GP formulation performs well and requires only a small fraction of queries. For the largest network, only $2.3\%$ of edges are queried, and for the smallest network, about $11\%$ of the edges are queried, in the worst case when using the goal programming ILP formulation.

\begin{table}[H]
\begin{center}
\caption{Results of ILP using Goal-Programming method}
\begin{tabular}{@{}lccccc@{}}
\toprule
\multicolumn{1}{c}{{\bf Network}} & \multicolumn{5}{c}{{\bf Number of measurements}}               \\ \midrule
\multicolumn{1}{c}{{\bf }}        & {\bf mean} & {\bf median} & {\bf mode} & {\bf max} & {\bf std} \\
{\bf Exnet}                       & 34.00      & 35           & 34         & 55        & 6.73      \\
{\bf CO. Springs}                 & 26.20      & 25           & 24         & 45        & 5.11      \\
{\bf Richmond}                    & 14.00      & 13           & 12         & 25        & 2.85      \\
{\bf Dtown}                       & 13.13      & 13           & 12         & 22        & 2.31      \\
{\bf Bangalore}                   & 12.10      & 12           & 12         & 17        & 1.28      \\ \bottomrule
\end{tabular}
\end{center}
\label{tab3:ILP_GP}
\end{table}

\begin{table}[H]
\begin{center}
\caption{Results of approximation algorithm}
\begin{tabular}{@{}lccccc@{}}
\toprule
\multicolumn{1}{c}{{\bf Network}} & \multicolumn{5}{c}{{\bf Number of measurements}}               \\ \midrule
\multicolumn{1}{c}{{\bf }}        & {\bf mean} & {\bf median} & {\bf mode} & {\bf max} & {\bf std} \\
{\bf Exnet}                       & 48.37      & 50           & 59         & 85        & 13.32     \\
{\bf CO. Springs}                 & 39.52      & 38           & 31         & 59        & 7.93      \\
{\bf Richmond}                    & 15.50      & 15           & 14         & 35        & 3.95      \\
{\bf Dtown}                       & 15.35      & 15           & 14         & 29        & 3.18      \\
{\bf Bangalore}                   & 12.00      & 12           & 11         & 18        & 1.65      \\ \bottomrule
\end{tabular}
\end{center}
\label{tab4:approx}
\end{table}

\newpage

\section{Conclusion}
An effective graph partitioning based protocol to locate leaky units in water distribution networks is proposed. The protocol involves solving a multi-objective optimization problem that approximately models hierarchical graph partitioning. It was observed that a goal programming formulation handles the multiple objectives in an effective manner, producing high quality solutions. An approximate partitioning algorithm inspired by spectral clustering was also presented, and the results discussed. The performance of the protocol and various formulations was elucidated through case studies on standard water distribution networks. It was observed that only a very small fraction of pipes need to be queried for flow measurements, in order to find the leak location.

\section{Extensions and Future Work}
In this section, we propose possible methods to avoid some assumptions made earlier. We also propose possible extensions and future work.

\subsection{Leaks in pipes}
As outlined earlier, extending the proposed protocol to include leak in pipes requires introducing more notations and modifying the protocol. For sake of brevity, we have presented this extension in the appendix. 

\subsection{Using existing sensors}
In the original problem formulation and protocol in Section~3, we assumed that pre-installed sensors are not available. Whenever a measurement was required, a query or act of measurement must be performed to obtain the flow rate. However, for well designed WDNs, some pipes would already be fitted with permanent sensors. This could be for DMA sectorization, or other monitoring requirements. In addition to sensors, we can also make use of valves by completely closing the valve through which we indirectly know that the flow rate in that pipe is zero. If such a disruptive method is not desirable, then the use of valves can be avoided. One method to incorporate these factors is to simply assign a very low querying cost to those pipes which have valves or sensors installed on them so that partitions containing them are favored over others. An extreme case of this is to simply remove those edges which have sensors on them from the network before running the partitioning algorithm and then use the appropriate flow rates when performing the water balance.

\subsection{Multiple leaks}
The proposed algorithm can be very naturally extended to cases where there are multiple leaks. In such a scenario, more than one sub-network would show an imbalance at some stage of the hierarchical partitioning exercise. After this point, we apply the same method to each of these sub-networks with imbalances. The only binding assumption in such a case is the absence of any material ingress - i.e. all the leaks are material losses out of the network, and water cannot enter the network through pipe ruptures.
\newline

\begin{protocol}[H]
\SetAlgoLined
{\bf Data:} Graph $\mathbf{G(N,E)}$ containing leaky node, $\delta$ (threshold) \\
{\bf Initialize:} Cost $\leftarrow$ 0;
    LeakySet  $\leftarrow$ $\lbrace \rbrace$ \\
{\bf Procedure:} (LeakySet, Cost) $\leftarrow$ FindLeak( $\mathbf{G}$, Cost, $\delta$, LeakySet ) \\
{\bf Result:} Leaky node(s) in LeakySet

\SetKwProg{Fn}{Function}{}{}
\Fn{\normalfont FindLeak( $\mathbf{G}$, RunningCost, $\delta$, LeakySet )}{
    \If{ $\text{size}(\mathbf{G}) > \delta$ }{
        $(\mathbf{S},\mathbf{\bar{S}}) \leftarrow $ partition $(\mathbf{G})$ \\
        Cost $\leftarrow$ RunningCost + $R(\mathbf{S},\mathbf{\bar{S}})$ \\
        \If{$\text{Leaky}(\mathbf{S})$}{
            (LeakySet, Cost) $\leftarrow$ FindLeak( $\mathbf{S}$, Cost, $\delta$, LeakySet )
        }
        \If{$\text{Leaky}(\mathbf{\bar{S}})$}{
            (LeakySet, Cost) $\leftarrow$ FindLeak( $\mathbf{\bar{S}}$, Cost, $\delta$, LeakySet )
        }
    }
    \Else{
        LeakySet $\leftarrow$ LeakySet + $\mathbf{N_G}$ \\
        Cost $\leftarrow$ RunningCost
    }
    {\bf Return:} (LeakySet, Cost)
}

\caption{Leak Detection Protocol}
\end{protocol}

\subsection{Different partitioning criteria}
In our work, we have tried to obtain partitions that are balanced in size of the sub-networks (measured in number of nodes). There are possibly alternate criteria for balanced partitions that take into account domain specific knowledge. For instance, if a probability distribution for leak occurrences in various nodes are available, we might want to obtain partitions that are balanced in this probability. This information could be obtained for instance through historical data or models utilizing network properties like pipe lengths, roughness factors etc. For instance, total length of pipe in a partition could be related to the probability of leak occurrence within the partition. It is easy to observe that node properties (like leak probability) can be easily incorporated into the ILP and Approximate algorithms. However, it is not trivial to partition based on edge attributes (like pipe length) which is a line of work we plan to pursue in the future.

\section*{Acknowledgments}
This work was partially supported by the Department of Science and Technology, India under the Water Technology Initiative (DST/TM/WTI/2K13/144) and the IIT Madras Interdisciplinary laboratory for data sciences (CSE/14-15/831/RFTP/BRAV).

\newpage

\appendix
\section*{Appendix}

In the main text, we presented the algorithm for finding leaks when they occur in nodes. However, in some cases, leaks may occur at any point along pipes as well. We now present an extension of the method for this case. We continue under the following assumptions:

\begin{enumerate}
    \itemsep0em
    \item The WDN is in steady state condition.
    \item The topology of the WDN (ie the graph representation) is known. 
    \item Flow meters can measure the flow and also detect the direction.
\end{enumerate}

\begin{figure}[b!]
\begin{center}
\includegraphics[width=0.95\textwidth]{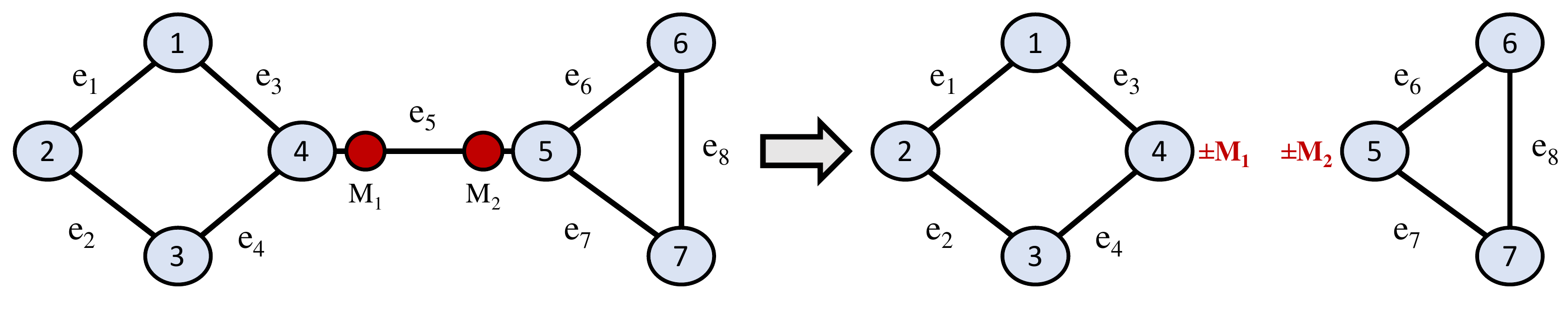}
\caption{Taking measurements close to nodes}
\label{fig:meas}
\end{center}
\end{figure}

We first present the idea for the simplistic case where there is a single leak and sensors are noiseless. Cosinder the graph $\mathbf{G}$ which contains the leak (either the full network, or network under consideration in some step of the recursive procedure). We consider a possible partition into $\mathbf{S}$ and $\mathbf{\bar{S}}$ by querying flows in cut$(\mathbf{S},\mathbf{\bar{S}})$. For the above scenario, a straightforward approach to querying a pipe is to measure the flows at both it's end points -- very close to the node, as shown in Fig.~7). If the flow at M$_1$ and M$_2$ do not match, it is clear the leak is in the pipe. However, if the flow rates at M$_1$ and M$_2$ are equal, then the leak is definitely not in this pipe. 
Following a similar procedure for all the pipes in cut$(\mathbf{S},\mathbf{\bar{S}})$, we can trace the leak to either $\mathbf{S}$ or $\mathbf{\bar{S}}$ exactly . In other words, the leaky node or pipe is within the partition. We would of course need to account for the flows by adding source or sink terms to the nodes on which the connecting pipes were incident. For example, in Fig.~\ref{fig:meas}, we need to add the flow rate in e$_5$ by adding a source or sink term at nodes 4 and 5, depending on the direction of flow. In this strategy, the cost will be twice the cut-cost.

\begin{figure}[H]
\begin{center}
\includegraphics[width=0.8\textwidth]{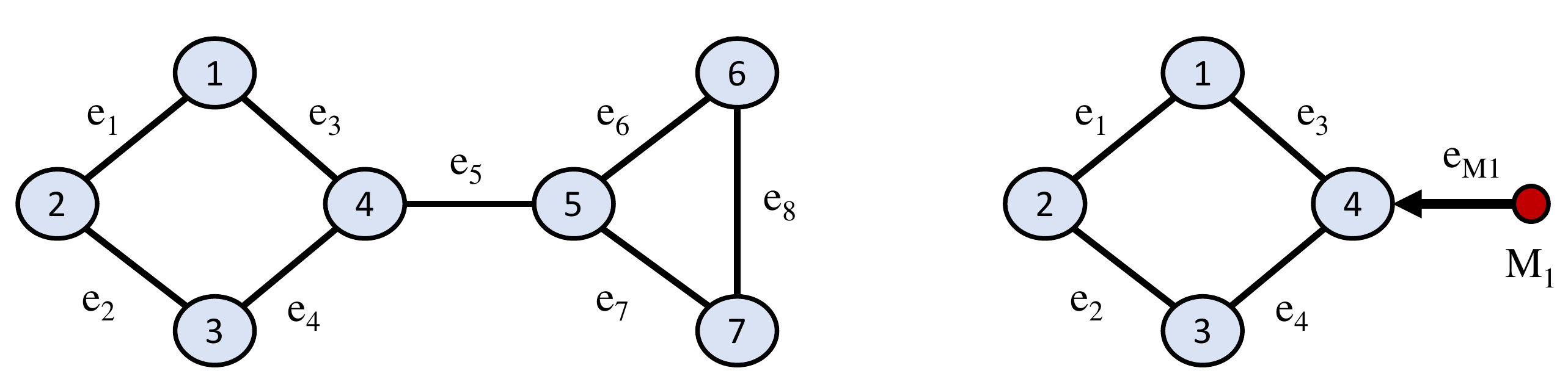}
\caption{Illustration of an artificial node $(M_1)$} 
\label{fig:intermittent}
\end{center}
\end{figure}

However, it is possible to reduce the cost with some modifications. When querying a pipe for the first time, rather than making two measurements, we can measure the flow at a single point close to the center. 
In this case however, the leak need not be in the interior of either partition. Since leaks can occur at any point on a pipe, the half-pipe segments of the crossing pipes (part of cut-set) could contain the leak. Thus we most modify our definition of partition to include these pipe segments as well (which are incident on only one node). 
We do this by introducing an {\it artificial} node at the point of measurement. Thus an edge between an actual node and artificial node represents a pipe segment. With this modification, the recursive procedure proposed in the main paper can be used. After many recursion steps, we may come to a stage where we need to query an edge between an actual node and artificial node. This amounts to measuring a pipe for the second time, where we already have one measurement for the pipe. In such a case, the second measurement is made close to the actual node. If this measurement does not match with the flow measurement obtained at the artificial node, then the pipe segment contains the leak and the process can be stopped. When the measurements match, we continue with the recursive procedure by eliminating the pipe segment. In this procedure, only a few pipes will be measured twice and, therefore, the cumulative number of measurements required will be lower. For illustration, consider the situation shown in Fig.~\ref{fig:intermittent}. An artificial node at point of measurement in added and the corresponding incidence matrix is:

{\small
$$ \mathbf{ J } = 
\begin{blockarray}{cccccc}
    \matindex{} & \matindex{e1} & \matindex{e2} & \matindex{e3} & \matindex{e4} & \matindex{eM1} \\
    \begin{block}{c(rrrrr)}
	\matindex{n1} & 1  & 0  & 1   & 0  & 0   \\
	\matindex{n2} & -1 & 1  & 0   & 0  & 0   \\
	\matindex{n3} & 0  & -1 & 0   & 1  & 0   \\
	\matindex{n4} & 0  & 0  & -1  & -1 & 1   \\
	\matindex{M1} & 0  & 0  & 0   & 0  & -1  \\
    \end{block}
\end{blockarray} $$
}

\end{document}